\documentclass[conference]{IEEEtran}
\usepackage[T1]{fontenc}
\usepackage{psfrag,amsmath,amssymb,color,cite, amsmath,graphicx}
\usepackage{stfloats}
\usepackage{mathtools}

\newcommand{\DeclareAutoPairedDelimiter}[3]{%
  \expandafter\DeclarePairedDelimiter\csname Auto\string#1\endcsname{#2}{#3}%
  \begingroup\edef\x{\endgroup
    \noexpand\DeclareRobustCommand{\noexpand#1}{%
      \expandafter\noexpand\csname Auto\string#1\endcsname*}}%
  \x}
\DeclareAutoPairedDelimiter\modulo{[}{]} 
\usepackage{fancyhdr,graphicx,amsmath,amssymb}
\usepackage[ruled,vlined]{algorithm2e}
\include{pythonlisting}
\usepackage{mathrsfs}
\usepackage{pifont}
\usepackage[colorinlistoftodos]{todonotes}

\newcommand{\beqn}{\begin{equation}}
\newcommand{\eeqn}{\end{equation}}
\newcommand{\beqa}{\begin{eqnarray}}
\newcommand{\eeqa}{\end{eqnarray}}
\newcommand{\beqas}{\begin{eqnarray*}}
\newcommand{\eeqas}{\end{eqnarray*}}

\presetkeys{todonotes}{color=green!30}{}
\usepackage{array}
\newcolumntype{P}[1]{>{\centering\arraybackslash}p{#1}}
\newcolumntype{M}[1]{>{\centering\arraybackslash}m{#1}}

\begin{document}

\title{Low Complexity Iterative Rake Detector for Orthogonal Time Frequency Space Modulation}
\author{\IEEEauthorblockN{Tharaj Thaj and Emanuele Viterbo\\}
\IEEEauthorblockA{ECSE Department, Monash University, Clayton, VIC 3800, Australia\\
Email: \{tharaj.thaj, emanuele.viterbo\}@monash.edu}}

\maketitle

\begin{abstract}
This paper presents a linear complexity iterative rake detector for the recently proposed orthogonal time frequency space (OTFS) modulation scheme. The basic idea is to extract and combine the received multipath components of the transmitted symbols in the delay-Doppler grid using linear diversity combining schemes like maximal ratio combining (MRC), equal gain combining  and selection combining to improve the SNR of the combined signal. We reformulate the OTFS input-output relation in the vector form by placing some null symbols in the delay-Doppler grid thereby exploiting the block circulant property of the channel matrix. Using the new input-output relation we propose a low complexity iterative detector based on the MRC scheme. The bit error rate (BER) performance of the proposed detector will be compared with the state of the art message passing detector and orthogonal frequency division multiplexing (OFDM) scheme employing a single tap minimum mean square error (MMSE) equalizer. We also show that the frame error rate (FER) performance of the MRC detector can be improved by employing error correcting codes operating in the form of a turbo decision feedback equalizer (DFE). 
\end{abstract}

\begin{IEEEkeywords} 
  OTFS, Detector, Decoder, Rake, Maximal Ratio Combining,  Delay--Doppler channel, turbo, DFE. 
\end{IEEEkeywords}
\section{Introduction}
Orthogonal time frequency and space (OTFS) is a new two dimensional (2D) modulation technique that transforms information symbols in the  delay-Doppler coordinate system to the familiar time-frequency domain \cite{Hadani} by spreading all the information carrying  symbols (e.g., QAM) over both time and frequency to achieve maximum diversity. As a result, the time-frequency selective channel is converted into an invariant, separable and orthogonal interaction, where all received QAM symbols experience the same localized impairment and all the delay-Doppler diversity branches are coherently combined.

OTFS can be imagined as a 2-D code division multiple access (CDMA) scheme where the information symbols are spread in both time and frequency as compared to either time or frequency as in the traditional CDMA systems \cite{Hadani}. A simple rake receiver in the case of direct sequence CDMA scheme in a multipath fading channel works by combining the delayed components or echoes of the transmitted symbols extracted using correlators matched to the respective orthogonal spread sequences (orthogonal time-frequency basis functions in the case of OTFS). Similarly, in the case of OTFS, the received delay and Doppler shifted components of the transmitted symbols in the OTFS grid can be extracted and combined using linear diversity combining techniques so as to maximize the SNR of the accumulated signal.

Diversity combining techniques are well studied in the literature starting from Brennan's paper on linear diversity combining \cite{MRC0}. Rake receivers for time domain combining using a variety of combining schemes like maximum ratio combining (MRC), equal gain combining (EGC) and selection combining (SC) are discussed in \cite{MRC01,MRC1}. Even though MRC is shown to work best when the branches are uncorrelated, it is  still shown to be optimal in the case of both correlated and uncorrelated branches as well as unequal noise and interference power in these branches \cite{MRC2,MRC3}. Moreover, iterative rake combining schemes and variants are shown to combat inter-symbol interference better and are well investigated in the literature for single and multi-carrier code division multiple access (CDMA) systems \cite{MRC5,MRC6}.

In this paper, we propose an iterative rake receiver for OTFS using the maximal ratio combining scheme. We start from the matrix input-output relation following \cite{Ravi2} and then group the delay-Doppler grid symbols into vectors according to their delay index and reformulate the input-output relation between the transmitted and received frames in terms of these transmitted and received vectors. By placing some null symbols in specific delay-Doppler grid locations we arrive at a reduced input-output relation, which is of the form that allows the use of the maximal ratio combining scheme to design a low complexity detector for OTFS. The number of null symbols, which can also be used as pilot symbols, needed for the proposed detection scheme is less than what is required for accurate channel estimation \cite{Ravi3} and so there is no additional utilization of resource or power for using these null symbols for detection. 

The rest of the paper is organized as follows. In Section II, we discuss the system model and derive the input-output relation in the vector form. In Section III, the proposed MRC based iterative rake detector and turbo-rake detector will be described. The simulation results are provided in Section IV along with some discussion on the complexity of the proposed algorithm in Section V. Section VI contains our concluding remarks.
\section{OTFS System Model}
\subsection{Notations}
The following notations will be followed in this paper; $a$, $\bf{a}$, ${\bf A}$ represents scalar, vector and matrix respectively. $a(n)$ represents the $n^{th}$ element of ${\bf a}$ and $a(m,n)$ represents the $(m,n)^{th}$ element of ${\bf A}$; ${\bf A}^H$, ${\bf A}^*$ and ${\bf A}^n$ represents the Hermitian transpose, complex conjugate and $n^{th}$ power of ${\bf A}$.  
The set of $M \times N$ dimensional matrices with complex entries in denoted by ${\mathbb{C}}^{N \times M}$. Let $\circledast$ represent circular convolution, $\circ$ the Hadamard product (the element wise multiplication) and $\oslash$  the Hadamard division and $|\mathcal{S}|$ the cardinality of the set $\mathcal{S}$. Let ${\bf F}_N$ and ${\bf F}_N^H$ be the N point DFT and IDFT matrices and ${\bf I}_M$ the $M \times M$ identity matrix. Let ${\bf 0}_N$ and ${\bf 1}_N$ denote a $N$ length column vector of zeros and ones respectively.  
\subsection{Transmitter and Receiver frames} 
The transmitter and receiver steps follows \cite{Ravi2,farhang}. Let {\bf X} and {\bf Y} be the transmitted and received two-dimensional symbols in the delay-Doppler grid. 
Let ${\bf x}_m$ and ${\bf y}_m$ be column vectors containing the symbols in the $m^{th}$ row of ${\bf X}$ and ${\bf Y}$ respectively: 
${\bf x}_m$ = $[{\bf X}(m,0), {\bf X}(m,1), \cdots, {\bf X}(m,N-1)]^T$ and ${\bf y}_m$ = $[{\bf Y}(m,0), {\bf Y}(m,1), \cdots, {\bf Y}(m,N-1)]^T$, where $m$ and $n$ denotes the delay and Doppler indices respectively, in the two-dimensional grid. 
We will be using this vector representation throughout the paper.
\subsection{Channel}
Consider a channel with $P$ propagation paths, where $h_i$, $l_i$,and $k_i$ are the complex {\em path gain, delay} and {\em Doppler shift index} associated with the $i^{th}$ path. 
The delay and Doppler-shift for the $i^{th}$ path is given by $\tau_i=\frac{l_i}{M\Delta f}$, $\nu_i=\frac{k_i}{NT}$.  
The total frame duration and bandwidth of the transmitted OTFS signal frame are $T_f=NT$ and $B = M \Delta f$, respectively. 
We consider the case where $T\Delta f=1$, i.e., the OTFS signal is critically sampled for any pulse shaping waveform. 
We assume that the maximum delay of the channel is $\tau_{\max} =l_{\max}T/M$ and that  the channel is under-spread, i.e., all $l_i\leq l_{\max}<M$ and $-N/2< k_i < N/2$. Since the number of channel coefficients, representing different scatterers, in the delay-Doppler domain is typically limited  the channel response has a sparse representation \cite{Hadani,Ravi2}:
\begin{equation} 
\label{eq:channel}
h(\tau, \nu) = \sum _{i=1}^{P} h_i \delta (\tau -\tau _i) \delta (\nu -\nu _i)  
\end{equation}
\subsection{Input-Output Relation}
Following \cite{Ravi2}, the input-output relation for the ideal pulse shaping waveform case can be written as a two dimensional circular convolution between ${\bf X}$ and the channel, i.e.,
\begin{align} {{\bf Y}} (m,n) = \sum _{i=1}^{P} h_i {{\bf X}} ([m-l_i]_M, [n-k_i]_{N})+w(m,n)\label{1} \end{align}
where $w(m,n)$ is iid AWGN noise with variance $\sigma_w^2$. 
In practical cases, the pulse shaping waveforms are not ideal, and the imperfect bi-orthogonality introduces extra phase shifts $\alpha_{i}(m,n)$ to each of the channel coefficients $h_i$. We assume a rectangular transmit and receive pulse shaping waveform as described in \cite{Ravi2,farhang}.
Following \cite{Ravi2}, the input-output relation for the rectangular pulse shaping waveform case (omitting the AWGN noise vector for brevity)  can be written as a two dimensional convolution in the form.     
    \begin{align} 
    {{\bf Y}} (m,n) = \sum _{i=1}^{P} h_i \alpha_i(m,n) {{\bf X}} ([m-l_i]_M, [n-k_i]_{N}) \label{2} 
    \end{align}
where  $z=e^{\frac{j2\pi}{MN}}$ and
    \begin{align} 
    \alpha _i(m,n) = \left. \left\lbrace \begin{array}{ll}e^{-j2\pi \frac{n}{N}} z^{k_i([m-l_i]_M)}, & \text{if } m < l_i\\ z^{k_i([m-l_i]_M)}, & \text{if } m \geq l_i\\ 0, & \text{otherwise.} \end{array}\right. \right.
    \label{3}
    \end{align}
We note that in this case we have a circular convolution
of ${\bf X}$ with a varying channel due to the phase terms in 
$\alpha _i(m,n)$.

First, following the notations described in the above subsection B, we can rewrite (\ref{2}) in  vector form by replacing ${{\bf Y}}(m, n) =  {{\bf y}}_{m}(n)$  and  ${{\bf X}}(m-l, [n-k]_N) =  {{\bf x}}_{m-l}([n-k]_N)$ as
\begin{align} {{\bf y}}_m(n) = \sum _{i=1}^{P} h_i \alpha_i(m,n) {{\bf x}}_{[m-l_i]_M}([n-k_i]_{N}) \label{vectorform} \end{align}
Equation (\ref{3}) gives two cases for the phase shifts introduced by the rectangular pulse shaping waveform. The first case, for phase shifts whit $m < l_i$, is dependent on both $m$ and $n$, whereas the second equation for $m \geq l_i$ depends only on $m$.
\begin{figure}
\centering
{\includegraphics[height=2.2in,width=3.3in]{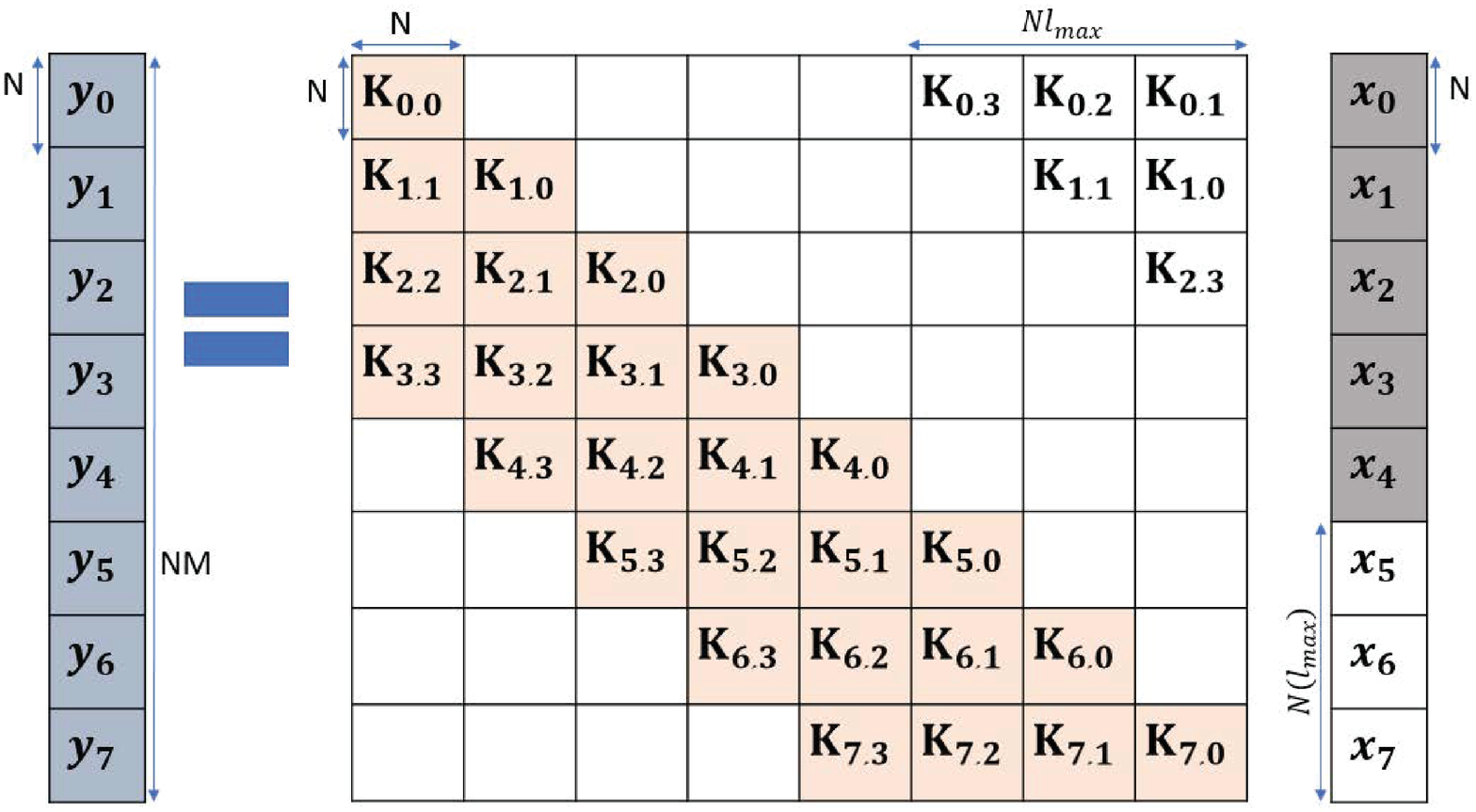}
 \vspace{-2mm}\caption{OTFS full channel matrix $({\bf H})$ after adding null symbols}
\label{mat2}}
\end{figure}
We may ignore the first case in (\ref{3}), which has a dependency on $n$ (Doppler index), by placing null symbol vectors ${\bf x}_m$ in  the last $l_{\max}$ rows of ${\bf X}$ such that, for all $l_i \leq l_{\max}$,
    \begin{align}
    h_i \alpha_i(m,n) {{\bf x}}_{[m-l_i]_M}([n-k_i]_{N})=0, \text{ if } m < l_i
    \label{condition1}
    \end{align} 
Hence, we can set
\begin{align}
{\bf x}_m(n)=0, \text{ if } m \geq M-l_{\max} \text{~and~} n=0,\ldots N-1 \label{cond2}
\end{align}

Fig. \ref{mat2} shows the $NM \times NM$ vectorized channel matrix ${\bf H}$ for OTFS for $N=M=8$ and $l_{max}=3$. As shown in Fig. \ref{mat2}, the transmitted and received symbol vectors, ${\bf x}_m$ and ${\bf y}_m$ respectively, are stacked in a column according to the respective delay indices ($m$). At the transmitter, the coloured vectors (${\bf x}_0,{\bf x}_1,{\bf x}_2,{\bf x}_3,{\bf x}_4$) denote valid symbol vectors and the non-coloured vectors (${\bf x}_5,{\bf x}_6,{\bf x}_7$) denote null symbol vectors (${\bf 0}_N$). 

The reduced phase correction equations for this null and valid data symbol placement now becomes
     \begin{align}
     \alpha _i^{\prime}(m) = \left. \left\lbrace \begin{array}{ll}  z^{k_i(m-l_i)}, & \text{if } m \geq l_{i}\\ 0, & \text{otherwise.} \end{array}\right. \right. \label{newphase} \end{align}
For $m=0,\cdots,M-1$ and $k$ = $0,\cdots,N-1$, let us define the vectors  ${\boldsymbol{\phi}}_{m} \in \mathbb{C}^{N \times 1}$, the phase correction vector containing the phases $\alpha _i^{\prime}(m)$ introduced by the non ideal pulse shaping waveform (rectangular in this case), with entries:
\begin{align} {\boldsymbol{\phi}}_{m}(k) = \left. \left\lbrace \begin{array}{ll}  z^{km}, & \text{if } 0 \leq k \leq N/2-1 \\ z^{-(N-k)m}, & \text{if } N/2 \leq k \leq N-1 \end{array}\right. \right.\label{phase} \end{align}
Let ${\boldsymbol{\nu}}_l \in \mathbb{C}^{N \times 1}$ be the channel Doppler spread vector at the  $l$-th delay tap  for ideal pulse shaping waveform, with entries:
   \begin{align} {\boldsymbol{\nu}}_l(k) = \left. \left\lbrace \begin{array}{ll}  h_i, & \text{if } l=l_i  \text{ and } k=[k_i]_N \\ 0, & \text{otherwise.} \end{array}\right. \right.
     \label{4} \end{align}
We can now rewrite (\ref{vectorform}), for $m < M-l_{\max}$, by replacing the channel coefficients $h_i$ and the reduced phase corrections $\alpha_i^{\prime}(m)$ with the channel Doppler spread vectors for ideal pulses ${{\boldsymbol{\nu}}_l}$ and phase correction vector ${\boldsymbol{\phi}}_{m}$ introduced by the rectangular pulses, 
    \begin{align}
    {{\bf y}}_m(n) = \sum _{l \in \mathcal{L}} \sum _{k = 0}^{N-1} {\boldsymbol{\nu}}_l(k) {\boldsymbol{\phi}}_{m-l}(k) {{\bf x}}_{m-l}( [n-k]_{N}) \label{5}
    \end{align}
where $\mathcal{L}=\{l_i \}$ is the set of unique delay tap indices among the  $P$ received paths in the delay-Doppler domain.

Now this can be written as the sum of one-dimensional circular convolutions between the vectors ${\boldsymbol{\nu}}_{m,l}$, ${\bf x}_{m-l} \in \mathbb{C}^{N \times 1}$, where ${\boldsymbol{\nu}}_{m,l}=[{\boldsymbol{\nu}}_{m,l}(0),{\boldsymbol{\nu}}_{m,l}(1),\cdots,{\boldsymbol{\nu}}_{m,l}(N-1)]$ 
\begin{align} 
{{\bf y}}_{m} = \sum _{l \in \mathcal{L}}{\boldsymbol{\nu}}_{m,l}\circledast{{\bf x}}_{m-l} \label{9} 
\end{align}
where
\begin{align} 
{\boldsymbol{\nu}}_{m,l}(k) = \left. \left\lbrace \begin{array}{ll}  {\boldsymbol{\nu}}_l(k){\boldsymbol{\phi}}_{m-l}(k), & \text{if } l \in \mathcal{L}, m \geq l\\ 0, & \text{otherwise.} \end{array}\right. \right. \label{10} 
\end{align}
Referring to the vectorized form shown in Fig. \ref{mat2}, we convert the circular convolution between two vectors into the product of a circulant matrix and a vector by defining ${\bf K}_{m,l} \in \mathbb{C}^{N \times N}$ to be a banded circulant matrix 
\begin{align*} {\bf K}_{m,l} & = 
  \text{circ}[{\boldsymbol{\nu}}_{m,{l}}(0),\cdots,{\boldsymbol{\nu}}_{m,{l_i}}(N-1)]\\[1ex]
  & = \left[\begin{array}{cccc} {{\boldsymbol{\nu}}_{m,{l}}}(0) & {{\boldsymbol{\nu}}_{m,{l}}}(N-1) & \cdots & {{\boldsymbol{\nu}}_{m,{l}}}(1)\\ {{\boldsymbol{\nu}}_{m,{l}}}(1) & {{\boldsymbol{\nu}}_{m,{l}}}(0) & \cdots & {{\boldsymbol{\nu}}_{m,{l}}}(2)\\ \vdots & \ddots & \ddots & \vdots \\ {{\boldsymbol{\nu}}_{m,{l}}}(N-1) & {{\boldsymbol{\nu}}_{m,{l}}}(N-2) & \cdots & {{\boldsymbol{\nu}}_{m,{l}}}(0) \end{array}\right] ~.
\end{align*}
From (\ref{4}) we note that the band width of each submatrix ${\bf K}_{m,l}$ of ${\bf H}$ is equal to the maximum Doppler spread $k_{\max}<N$  and the full channel matrix ${\bf H}$ has a band width equal to $N(l_{max}+1)$. We can then write (\ref{9}) as 
     \begin{align} {{\bf y}}_m = \sum _{l \in \mathcal{L}} {\bf K}_{m,l}\cdot{{\bf x}}_{m-l} \label{11} 
     \end{align}
Note that ${\bf K}_{m,l}$ can be considered as the time-varying Doppler spread matrix at the delay tap with index $l$. Now (\ref{9}) and (\ref{11}) gives us a very simple equation relating the transmitted and received symbol vectors that we defined at the start of this section. This is a much more compact form, compared to the input-output relation we began with. The vector relations shows how the symbol vector transmitted at delay index $m-l$ is impaired by the channel Doppler spread  vector ${\boldsymbol{\nu}}_{m,{l}}$ (or matrix ${{\bf K}}_{m,{l}}$) at the delay tap with index $l$.
\section{Low Complexity Iterative Rake Detector}\label{sec1}
We can think of the proposed MRC decoder as the maximal ratio combining of the channel impaired signal components received at  $L=|\mathcal{L}|\leq P$ different delay branches in the delay-Doppler grid analogous to the CDMA rake receiver. 
The SNR of the  received signal components of a transmitted symbol vector ${\bf x_m}$ in each of these branches are unequal and depends on the channel response. 
The optimal MRC weights in this case are discussed in \cite{MRC3}. In our proposed detector, we iteratively cancel inter-symbol interference in the branches we have selected for combining, so as to maximize the signal to noise ratio at the output of the MRC. 
      
We have the input output relation between the transmitted and received symbol vectors ${\bf x}_m$ and ${\bf y}_m$ given by
\begin{equation}{\bf y}_m={\sum_{l \in \mathcal{L}}{\bf K}_{m,l}\cdot{\bf x}_{m-l}} + {\bf w}_m \label{io}\end{equation}
 where ${\bf w}_m$ is iid AWGN noise with variance $\sigma_n^2$. Due to the inter-symbol interference caused by delay spread ($l_{\max} \Delta\tau$), all vectors ${\bf x}_m$ have a signal component in $L$ received symbol vectors ${\bf y}_{m+l}$ where $l \in \mathcal{L}$ (\ref{io}). Let ${\bf b}_m^{l} \in \mathbb{C}^{N \times 1}$ be the channel impaired signal component of ${\bf x}_m$ in the received vector at delay index $m+l$ (${\bf y}_{m+l}$) after removing the interference of the other transmitted symbol vectors ${\bf x}_{k \neq m}$. Assuming we have the estimates of symbol vectors ${\bf x}_m$ from previous iterations, we can then write ${\bf b}_m^{l}$ for $l\in \mathcal{L}$  as
        \begin{equation}
        {\bf b}_m^{l} ={\bf y}_{m+l}-{\sum_{l^{\prime}\in \mathcal{L},l^{\prime}\neq l}{\bf K}_{m+l,l^{\prime}}}\cdot\hat{\bf x}_{m+l-l^{\prime}}
        \label{highcomp}\end{equation}
Then from (\ref{io}) and (\ref{highcomp}) for $l\in \mathcal{L}$, we have $L$ equations for the symbol vector estimates $\hat{\bf x}_{m}^{(l)}$ given as
         \begin{equation}
        {\bf b}_m^{l} ={{\bf K}_{m+l,l}}\cdot{\hat{\bf x}_m}+{\bf w}_{m+l}
        \label{dfe}\end{equation}
In our proposed scheme, instead of estimating the transmitted symbol vector $\hat{\bf x}_m$ separately from each of the $L$ equations in (\ref{dfe}), we maximal ratio combine the estimates ${\bf b}_m^{l}$ (\ref{Rm}) and then decode vectors $\hat{\bf x}_m$ 
symbol-by-symbol by using the ML criterion as given below in (\ref{ML}). Let us define 
\begin{equation}
    {\bf R}_{m}={\sum_{l \in \mathcal{L}}{\bf K}_{m+l,l}^H\cdot{{\bf K}_{m+l,l}}}\label{denom}
\end{equation}
\begin{equation}
    {\bf g}_m=\sum_{l \in \mathcal{L}}{\bf K}_{m+l,l}^H\cdot{\bf b}_m^{l}\label{num}
\end{equation}

Then the output of the maximal ratio combiner, ${\bf c}_m \in \mathcal{C}^{N \times 1}$, is given by 
 \begin{equation}{\bf c}_m ={\bf R_{m}}^{-1}\cdot{\bf g}_m \label{Rm} \end{equation}
\begin{equation} {\bf \hat{x}}_m(n)=\arg \min _{a_j\in \mathcal {Q}} \left |{a_j-{\bf c}_m(n)}\right |. \label{ML}\end{equation}
where $a_j$ is an element from the set of transmitted QAM alphabet $\mathcal{Q}$ with  $j=1,\cdots,|\mathcal{Q}|$ and $n=0,\cdots,N-1$.
Once we update the estimate $\hat{\bf x}_m$, we increment $m$ and repeat the same to estimate all  $M'=M-l_{max}$ information symbol vectors ${\bf \hat{x}}_m$  using the updated estimates of the previous decoded symbol vectors in the form of a decision feedback equalizer (DFE).
        
\begin{algorithm}
\SetAlgoLined
\For{$m=0:M'-1$}{
        ${\bf R}_{m}={\sum_{l \in \mathcal{L}}{\bf K}_{m+l,l}^H{{\bf K}_{m+l,l}}}$\\}
    \For{iteration=1:max}{
    ${\bf g}_m={\bf 0}_N$\\
    \For{$m=0:M'-1$}{
       \For{$l \in \mathcal{L}$}{
            ${\bf b}_m^{l} =({\bf y}_{m+l}-{\sum_{l^{\prime} \neq l}{\bf K}_{m+l,l^{\prime}}}\cdot\hat{\bf x}_{m+l-l^{\prime}})$\\
            ${\bf g}_m={\bf g}_m+{\bf K}_{m+l,l}^H\cdot{\bf b}_m^{l}$\
    }
    ${{\bf c}}_m ={\bf R}_m^{-1}\cdot{\bf g}_m$\\
    ${\bf \hat{x}}_m(n)={\arg \min} _{a_j\in \mathcal {Q}} \left |{a_j-{{\bf c}}_m(n)}\right |$\
    }}  
\label{algo1}\caption{MRC Rake Detector}
\end{algorithm}
It can be seen from  (\ref{denom}) that ${\bf R}_m$ is the sum of product of circulant matrices ${\bf K}_{m,l}$ and hence a circulant matrix, which can be computed in the Fourier domain in $M'NL$ computations. In (\ref{highcomp}), for each symbol vector ${\bf x}_m$, we need to compute $L$ vectors ${\bf b}_m^{l}$. This operation requires $L(L-1)$ products between circulant matrices ${\bf K}_{m,l}$ and symbol vectors ${\bf x}_{m-l}$. We can take advantage of the redundant summation operations to reduce the complexity of (\ref{highcomp}).
By defining 
 \begin{equation}
     {\bf {\hat y}}_{m+l}={\sum_{l^{\prime} \in \mathcal{L}}{\bf K}_{m+l,l^{\prime}}}\cdot\hat{\bf x}_{m+l-l^{\prime}}
\label{yhat}\end{equation}
we can  rewrite (\ref{highcomp}) as 
\begin{equation}
        {\bf b}_m^{l} ={\bf y}_{m+l}-{\bf {\hat y}}_{m+l}+{\bf K}_{m+l,l}\cdot\hat{\bf x}_{m}
        \label{lowcomp12}\end{equation}
The $L$ vectors ${\bf {\hat y}}_{m+l}$ in (\ref{yhat}) requires computation of $L^2$ matrix-vector products. Let ${\bf x}_m^{(i)}$ be the estimate of ${\bf x}_m$ computed in the $i^{th}$ iteration. Then in the $i+1^{th}$ iteration for every $m=0, \cdots, M-l_{max}-1$ and $l \in \mathcal{L}$, instead of computing the $L$ vectors ${\bf {\hat y}}_{m+l}$ again with the latest estimates $({\bf x}_m^{(i+1)})$ using (\ref{yhat}), we can simply update the vectors ${\bf {\hat y}}_{m+l}$ and ${\bf b}_m^l$ as follows 
\begin{equation}
        {\bf {\hat y}}_{m+l}={\bf {\hat y}}_{m+l}+{\bf K}_{m+l,l}\cdot({\hat{\bf x}}_{m}^{(i+1)}-{\hat{\bf x}}_{m}^{(i)})
        \label{lowcomp2}\end{equation}
\begin{equation}
        {\bf b}_m^{l} ={\bf y}_{m+l}-{\bf {\hat y}}_{m+l}+{\bf K}_{m+l,l}\cdot\hat{\bf x}_{m}^{(i+1)}
        \label{lowcomp3}\end{equation}
Each of (\ref{lowcomp2})  and (\ref{lowcomp3}) for all $l \in \mathcal{L}$ requires $L$ matrix-vector product computations per symbol vector ${\bf x}_m$. If we compute ${\bf {\hat y}}_{m+l}$ and store it, then only (\ref{lowcomp2})  and (\ref{lowcomp3}) needs to be calculated in every iteration. The overall number of matrix-vector products for estimating vectors ${\bf b}_m^l$ for each $m$ is then reduced from $L(L-1)$ in (\ref{highcomp}) to $2L$ in (\ref{lowcomp2}), (\ref{lowcomp3}). The vectors ${\bf g}_m$ and ${\bf c}_m$ then together requires $L+1$ matrix-vector products per symbol vector ${\bf x}_m$ per iteration. 


The matrix-vector products in Algorithm \ref{algo1} are products between circulant matrices ${\bf K}_{m,l} \in {\mathbb C}^{N \times N}$ and column vectors ${\bf x}_{m}\in {\mathbb C}^{N \times 1}$ which can be converted to element-wise product of vectors in the Fourier domain with a complexity of $N$ complex multiplications. Overall complexity per iteration for calculating ${\bf b}_m^l$, ${\bf g}_m$ and ${\bf c}_m$ for all symbol vectors is then $M'(3L+1)N$ complex multiplication. The redundant summations and FFT computations can be avoided by storing the Fourier transform of the first column of all $M'L$ circulant matrices ${\bf K}_{m,l}$, $M'$ vectors ${\bf x}_m$ and $M'L$ vectors ${\bf {\hat y}}_{m+l}$ in (\ref{yhat}), and then operating in the Fourier domain. 
\subsection{Low Complexity Initial Estimate}
In Algorithm \ref{algo1}, we initially assume that all the alphabets of the QAM modulation set $\mathcal{Q}$ are equally likely and hence we initialize $\hat{{\bf x}}_m ={\bf 0}_N$, for all $m$. Even though the MRC detector complexity per iteration is of the order $O(NML)$, the overall complexity scales linearly with the number of detector iterations needed to converge. 

However, a low complexity initial estimate of the OTFS symbols may reduce the required number of MRC detector iterations and hence the overall complexity. A single tap equalizer assuming ideal pulse shaping waveform in the time-frequency domain can provide a low complexity rough initial estimate of the OTFS symbols. 

Define ${\bf H}_{dd}(m,n) \in \mathbb{C}^{M \times N}$, the delay-Doppler domain channel impulse response matrix for the ideal pulse shaping waveform case, 
\begin{align*} {\bf H}_{dd}(m,n) = \left. \left\lbrace \begin{array}{ll} h_i, & \text{if } m=l_i, n=[k_i]_N\\0, & \text{otherwise } \end{array}\right. \right. \end{align*}
The corresponding time-frequency channel response for the ideal pulse shaping waveform is obtained by an ISFFT operation on the delay-Doppler channel as
\begin{equation}{\bf H}_{tf}={\bf F}_M{\bf H}_{dd}{\bf F}^H_N\end{equation}
Similarly  the received time-frequency samples can be obtained by the ISFFT operation on the received delay-Doppler domain samples as
\begin{equation}{\bf Y}_{tf}={\bf F}_M{\bf Y}{\bf F}^H_N\end{equation}
Since in the ideal pulse shaping waveform case, circular convolution of the channel and transmitted symbols in the delay-Doppler domain transforms to element-wise product in the time-frequency domain, we estimate the transmitted samples in the time-frequency domain by a single tap  minimum mean square error (MMSE) equalizer 
\begin{equation}
    {\bf \hat{X}}_{tf}=({\bf H}_{t,f}^{\ast}\circ{\bf Y}_{t,f}){\oslash}(\lvert {\bf H}_{t,f}\rvert^{2}+\sigma_w^2)
    \label{20}
\end{equation}
where $\oslash$ represents the Hadamard division (element wise division) and superscript $\ast$ denotes the complex conjugate.

The delay-Doppler domain estimate the OTFS symbols can then be obtained by the SFFT operation on the time-frequency domain estimates as 
    \begin{equation}
     {\bf \hat{X}}={\bf F}^H_M{\bf \hat{X}}_{tf}{\bf F}_N
    \end{equation}
Then the initial estimate of the symbol vectors are simply ${\bf \hat{x}}_m^{(0)}$ = [${\bf \hat{X}}(m,0)$, ${\bf \hat{X}}(m,1)$ \dots  ${\bf \hat{X}}(m,N-1)]^T$.
\subsection{Turbo Rake Detector}
The frame error rate (FER) performance of the detector can be improved by employing an error control code. The encoded bits are random interleaved in the frame so as to extract maximum time and frequency diversity. 

\begin{figure}
\centering
{\includegraphics[height=2in,width=3.1in]{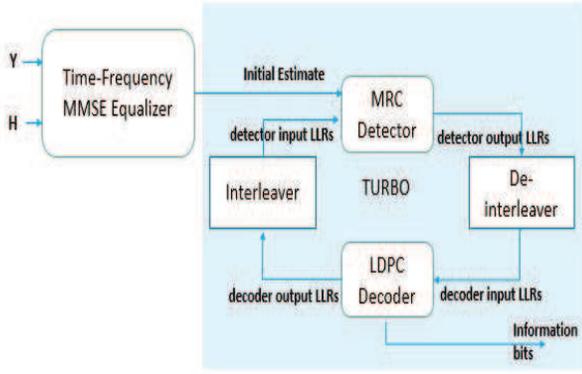}
 \vspace{-3mm}\caption{Turbo-MRC Operation}
\label{turbo}}
\end{figure}

The turbo decoder principle as shown in Fig. \ref{turbo} can be used to further improve the FER performance. The detector output bit log likelihood ratios (LLR) after random de-interleaving is fed to the LDPC decoder. The output bit LLRs  from the LDPC decoder after interleaving is then fed back to the MRC detector and the process repeats. Inside the MRC detector a hard decision is taken on the input LLRs from the LDPC decoder to get the estimates of ${\bf x}_m$. Overall, one turbo iteration involves one iteration of MRC detector, de-interleaver, LDPC decoder and interleaver. 
\section{Simulation Results and Discussion}
For simulations we use an OTFS frame with $N=128$ and $M=512$ and sub-carrier spacing of 15 KHz. The maximum delay spread (in terms of integer taps) is taken to be 32 ($l_{max}=31$) which is approximately 4 ${\mu}s$. The channel delay model is generated according to the standard Extended Vehicular A (EVA) model (speed = 120 km/hr) with the Doppler shift for the $i^{th}$ path generated from a uniform distribution $U(0,\nu_{max})$, where $\nu_{max}$ is the maximum Doppler shift. The EVA channel power delay profile is given by [0, -1.5, -1.4, -3.6, -0.6, -9.1, -7.0, -12.0, -16.9] dB with excess tap delays [0, 30, 150, 310, 370, 710, 1090, 1730, 2510] ns \cite{EVA}.   We consider one Doppler shifted path per delay tap with $L=9, l_{max}=32$ and $k_{max}=16$ and  in the simulations.
  
\begin{figure}
    \centering
    {\includegraphics[height=2.1in,width=3.2in]{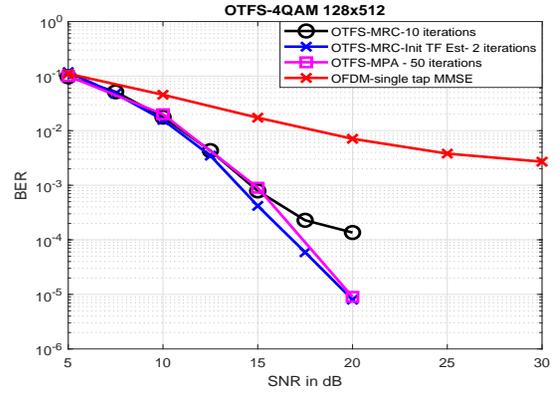}
     \vspace{-3mm}\caption{Uncoded 4-QAM BER Plot : MRC vs MPA vs MMSE-OFDM}
    \label{ber_4qam}}
    \end{figure}

\begin{figure}
    \centering
    {\includegraphics[height=2.1in,width=3.2in]{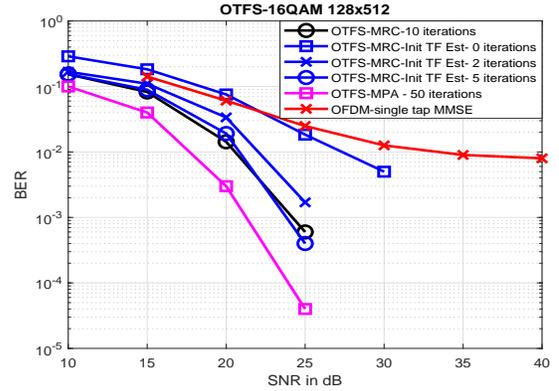}
     \vspace{-3mm}\caption{Uncoded 16-QAM BER Plot : MRC vs MPA vs MMSE-OFDM}
    \label{ber_16qam}}
    \end{figure}

\begin{figure}
    \centering
    {\includegraphics[height=2.1in,width=3.2in]{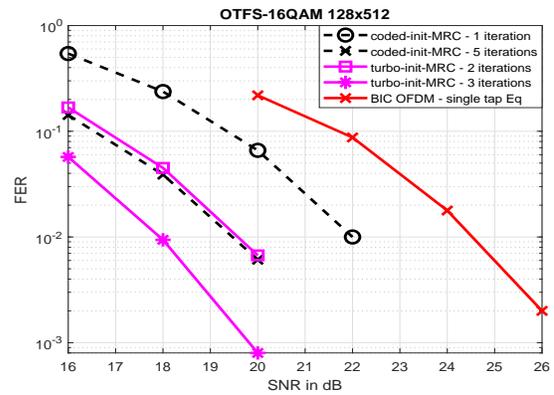}
    \vspace{-3mm}\caption{  Coded 16-QAM FER Plot: MRC vs BIC-OFDM-MMSE}
    \label{turboplot}}
    \end{figure}

Fig. \ref{ber_4qam} shows the BER plot for the MRC detector for 4-QAM modulated OTFS waveform with 10 iterations comparing it with the state of the art message passing algorithm (MPA) described in \cite{Ravi,Ravi1} (labelled as OTFS-MPA in Fig. \ref{ber_4qam} and \ref{ber_16qam}) with 50 maximum iterations (the message passing algorithm has a stopping criteria based on convergence of the estimated symbol probabilities) and the OFDM single tap MMSE equalizer. In Fig. \ref{ber_4qam}, we show the performance of the MRC detector with the initial estimate obtained using a single-tap MMSE equalizer in the time-frequency domain. As we can see, there is a 1 dB gain at a BER of $10^{-3}$ with just 2 iterations of the MRC detector with the initial estimate (MRC-Init-TF Est).\\
Fig. \ref{ber_16qam} shows the BER plot for the MRC detector for 16-QAM modulation with 10 iterations. Using the initial estimate from the single tap equalizer in the time-frequency domain, the number of iterations can be reduced. Only 5 iterations are needed to match the BER performance of plain MRC detector with 10 iterations, which approximately halves the overall complexity. The 16-QAM BER performance is compared with the OFDM scheme, and we see that the initial time-frequency estimate itself (curve corresponding to MRC-Init TF Est for 0 iterations in the plot) performs better than the single tap OFDM scheme. The error performance is further improved by the MRC detector iterations.\\
Fig. \ref{turboplot} shows the frame error performance of a turbo and plain coded MRC detector with the half rate LDPC codeword of length 4096 bits. The cases with and without the turbo iterations is plotted. Please note that coded and turbo MRC detector is the same for 1 iteration. It can be observed that just 1 iteration of plain coded MRC detector (coded-init-MRC) is required to achieve better error performance than the bit interleaved coded MMSE OFDM. Moreover, we can gain further by turbo operation as explained in the previous section. It can be seen that with 2 iterations of turbo MRC detector (turbo-init-MRC) we can achieve the same performance as a plain coded MRC detector with 5 iterations. The overall detector complexity in the form of required iterations is significantly reduced by using the initial estimates from the time-frequency single tap equalizer along with turbo operation.
\section{Detector Complexity}
 The actual overall complexity (in terms of complex multiplications), including initial computations and Fourier domain transformations as discussed at the end of Section \ref{sec1},  is 
  \[
 \overbrace{NM'S(3L+1)}^{(1)}+\overbrace{NM'L^2}^{(2)}
 \]
\[
 \overbrace{NM'(2L+1)\log_2(N)}^{(3)}+\overbrace{NM[3+3\log_2(NM)]}^{(4)}
 \]
where $S$ is the number of MRC detector iterations. The term (1) includes the iterative computations inside detector (calculating ${\bf b}_m^l$, ${\bf g}_m$ and ${\bf c}_m$) in Fourier domain and term (2) is for calculating the initial $M'L$ vectors ${\bf {\hat y}}_{m}$ in (\ref{yhat}) and $M'$ vectors ${\bf R}_m$. The term (3) includes computing the FFT of the first column of the $M'L$ circulant matrices ${\bf K}_{m,l}$,\footnote{Operations in (3) are part of the channel estimation process and can be simplified by using the time domain received pilot samples.} the $M'L$ vectors ${\bf {\hat y}}_{m}$ and $M'$ vectors ${\hat {\bf x}}_m$ and term (4) is for computing the low complexity initial time-frequency estimate ${\hat {\bf x}}_m^{(0)}$ (\ref{20}).

The linear complexity detectors currently available in the literature for OTFS \cite{Ravi,kgp} with non ideal pulse shaping waveform (rectangular) are still not of lower enough complexity for practical applications. The complexity of MPA detector scales with alphabet size $|Q|$ and has a complexity of the order of $O(SNMP|Q|)$ \cite{Ravi}. The storage requirement for the MRC detector is in the order of $O(NML)$, whereas for MPA it is $O(NMP|Q|)$ \cite{Ravi}. The detector proposed in \cite{kgp} even though is a non iterative detector has a computational complexity of $O(MNk_{\max}P^2)$ where $k_{\max}$ is the maximum Doppler spread, whereas our proposed detector has a complexity of $O(SMNL)$ where $L \leq P$.\footnote{By selecting only the dominant paths, $L$ can be reduced.} 

\section{Conclusion}
We reformulated the OTFS input-output relation and proposed a {\em linear complexity} iterative rake detector algorithm for OTFS modulation based on the maximal ratio combining scheme. We show that the MRC detector can achieve similar BER performance as compared to MPA detector but with lower complexity and storage requirements. The required number of iterations and hence complexity can be reduced by employing a low complexity single tap MMSE equalizer in the time-frequency domain to get an initial estimate of the OTFS symbols. The MRC detector performance can be further improved with the aid of error control codes and through turbo iterations.

\end{document}